\documentclass[journal,letterpaper]{IEEEtran_nssmic}

\usepackage{cite}      

\usepackage{graphicx}  

\usepackage{url}       

\usepackage{amsmath}   
\begin{document}
%
\title{High Efficiency Detection of Argon Scintillation Light of 128nm Using LAAPDs}
%
%
\author{\authorblockN{Rico~Chandrasekharan\authorrefmark{1}\authorrefmark{3},
Andreas~Knecht\authorrefmark{1}, Marcello~Messina\authorrefmark{1},
Christian~Regenfus\authorrefmark{2}, and
Andr\'e~Rubbia\authorrefmark{1}}\\
\authorblockA{\authorrefmark{1} ETH Z\"urich, Institut f\"ur Teilchenphysik, CH-8093 Z\"urich, Switzerland\\
\authorrefmark{2} University of Z\"urich, Physik Institut, Winterthurerstr. 190, CH-8057 Z\"urich, Switzerland\\
\authorrefmark{3} Email: rico.chandrasekharan@cern.ch, Telephone: ++41 22 767 1474, Fax:  ++41 22 767 1411 \\
}}

\maketitle

\begin{abstract}
The possibility of efficient collection and detection of vacuum
ultraviolet light as emitted by argon, krypton, and xenon gas is
studied. Absolute quantum efficiencies of large area avalanche
photodiodes (LAAPDs) are derived at these wavelengths. VUV light of
wavelengths down to the 128nm of Ar emission is shown to be detectable
with silicon avalanche photodiodes at quantum efficiencies above
42\%. Flexible Mylar foil overcoated with Al+MgF$_2$ is measured to
have a specular reflectivity of $\sim$91\% at argon emission
wavelength. Low-pressure argon gas is shown to emit significant
amounts of non-UV radiation. The average energy expenditure for the
creation of non-UV photons in argon gas at this pressure is measured
to be below 378 eV.

\end{abstract}


\section{Introduction}

\PARstart{T}{he} ArDM detector currently under construction is a WIMP
detector prototype based on argon bi-phase technology, see
\cite{Rubbia:2005ge}. In this detector, the detection of the 128 nm argon
scintillation light as well as a localized charge readout allow
discrimination against gamma/beta backgrounds. A major goal of the
project is the proof of scalability of such technology to large target
masses. ArDM is aiming to be among the first large direct dark matter
detectors, with a target mass of the order of one metric ton.

In the last two years extensive R$\&$D has been conducted towards the
achievement of this goal. This included investigation of the
possibility to reflect VUV photons on Al coated mylar foils as a means
of light collection. For light detection purposes, the feasibility of
using large area avalanche photodiodes (LAAPD) for light read-out was
analyzed. This talk presents the results of the measurement of the
quantum efficiency of such devices.

Noble gases are known to provide scintillation light with high
yield,up to about 50 photons per keV electron equivalent. Detection of
this light proves fundamental in many applications in which noble gas
is used as medium. Difficulties arise from the fact that noble gas
emission is peaked in the vacuum ultraviolet (VUV) range. To detect
such light, wavelength shifter-coated photomultiplier tubes (PMTs)
have often been used, resulting in a low global quantum efficiency,
typically $\le10\%$. There is interest in knowing if LAAPDs could be a
viable alternative, in particular in applications where the
radiopurity is a concern.

The presented results show that it is in principle possible both to
collect VUV light using mirrors, as well as to detect gas
scintillation from Kr, Xe and Ar with APDs at a quantum efficiency
significantly higher than with PMTs. The issues related to the signal
to noise of APDs, to the parallel operation of a large number of APDs
to increase the sensitive area, and the mechanical problems at
potential cryogenic temperatures are not addressed in this paper.

 

\section{Experimental Setup}

\begin{figure}
\centering
\includegraphics[width=2.5in]{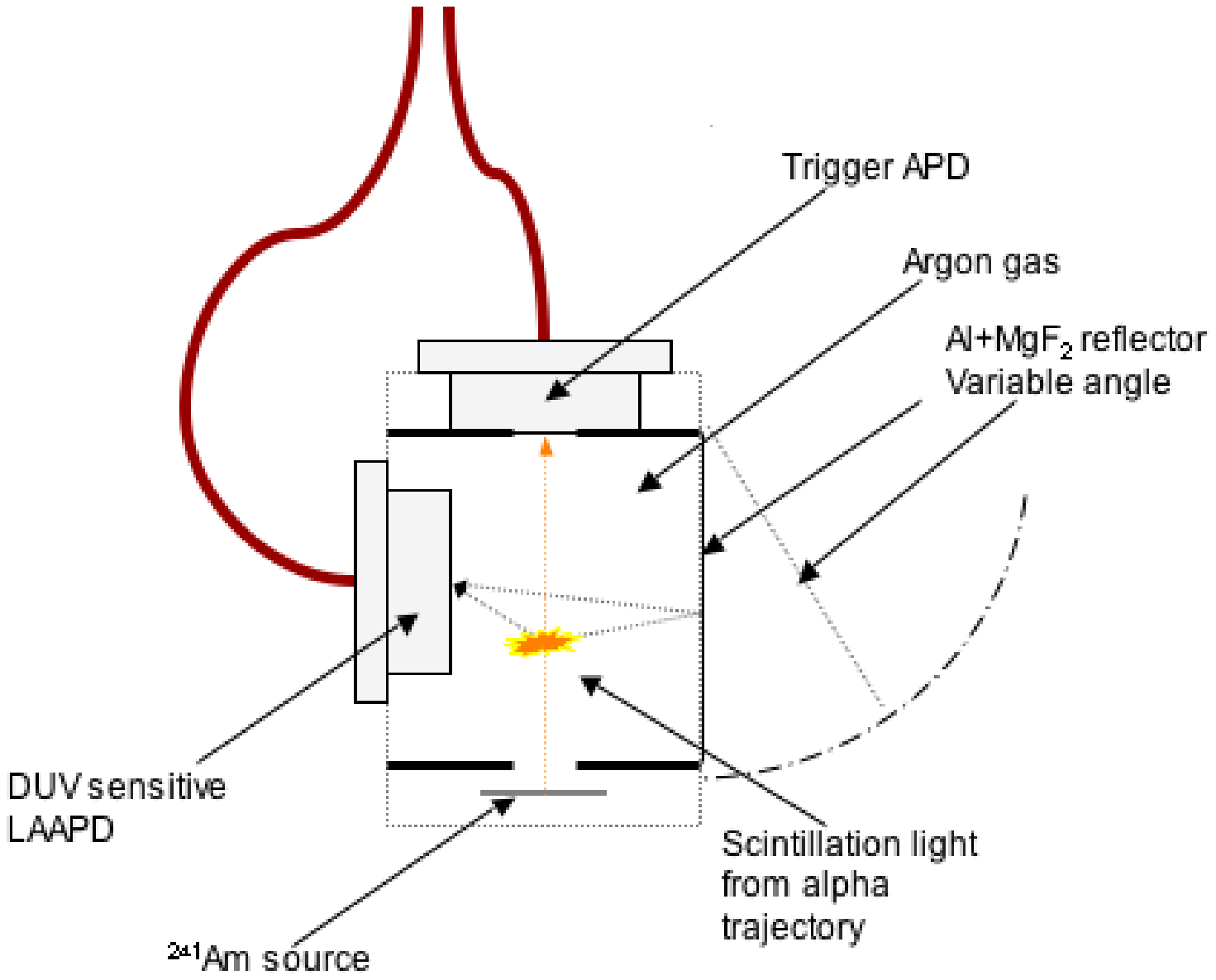}
\caption{\label{fig:setup} Alpha particles pass from the source to an
APD serving as trigger. A laterally mounted APD measures the
scintillation light emitted along the trajectory. A VUV
reflecting mirror may be placed at a desired angle, enhancing
the signal.}
\end{figure}

In our experimental setup (See Figure~\ref{fig:setup}), we detect
argon gas scintillation light with an LAAPD mounted on an axis
perpendicular to the trajectory of alpha particles. The alpha
particles pass from the open $^{241}$Am source to a second APD
employed as a trigger. The setup for quantum efficiency measurements is
described in more detail in~\cite{rico}.

The APDs are Advanced Photonix LAAPDs with an active diameter of 16mm
\cite{advphot}. For scintillation light detection a windowless,
'DUV-enhanced' device was used primarily, cross-checked against a
windowless 'Red/IR-enhanced' device.

At the gas pressures used, the decay time of the argon scintillation
signal is of the order of several microseconds~\cite{moerman}. Long
shaping times allow full integration of the complete signal.

Measuring the current flowing through the APD allows the gain to be
monitored according to the method described in \cite{Karar:1999pg},
where a LED situated inside the dewar serves as a continuous light
source.

For the reflectivity measurements presented in Section~\ref{refl}, a
VUV reflecting mirror (also shown in Figure~\ref{fig:setup}) may be
placed at a desired angle, enhancing the signal. Via a mechanical
feed-through, the mirror angle may be set to values between $\phi=0$
and $\phi=90$ degrees, where $\phi$ is defined as the angle between
the mirror plane and the alpha trajectory. The setup for reflectivity
measurements is described in more detail in~\cite{knecht}.

\section{Simulation of Photon Yield}

The gases used in this work- Ar, Kr, and Xe- emit photons around
128nm, 150nm, and 175nm, respectively. The VUV photon production
mechanism in argon and other noble gases is known, see for
example~\cite{miyajima} and citations therein/thereof. Accordingly,
the average energy expenditure for the emission of a VUV scintillation
photon in pure argon gas is $W_\gamma^{VUV}=$ 67.9~eV. Similarly, for
krypton and xenon, $W_\gamma^{VUV}=$ 61.2~eV and 55.9~eV,
respectively (See Appendix).

The $\alpha$'s trajectory in our setup is well defined due to
collimation. Given the gas temperature and pressure, the energy loss
of the $\alpha$ along its trajectory can then
computed~\cite{NIST}. The number of VUV photons created and emitted
into the solid angle subtended by the LAAPD can be simulated for an
$\alpha$ following its trajectory. In the following, this number is
referred to as $N_\gamma^{sim}$.

For the reflectivity measurements of Section~\ref{refl}, the
simulation additionally computes the number of photons reaching the APD
indirectly through specular reflection by the mirror placed at a
variable angle $\phi$. This value is referred to as
$N_\gamma^{sim}(\phi)$.

The simulated ratio of indirect i.e. reflected to direct light is
\begin{equation}
N_\gamma^{sim}(\phi)/N_\gamma^{sim}.
\end{equation}
This can then be compared to the measured values.

\section{Quantum Efficiency Measurement}

A number of measurements were performed in argon, krypton, and xenon
gas. The gas pressures were chosen to ensure the alphas depositing
sufficient energy in the trigger APD.

\begin{table}
\renewcommand{\arraystretch}{1.3}
\caption{The gain-independent signal normalized to the expected number
of VUV photons in argon. Measurements performed using 'DUV-enhanced'
LAAPD. All data was acquired using a shaping time of 10 $\mu$s. The
last line gives the average value of the measurements.}
\label{tab:ar}
\centering
\begin{tabular}{|c|c|c|c|c|c|}
\hline
 $p_{Ar}$  & $T_{Ar}$  & Gain & $Q_s$  &  $N_\gamma ^{sim}$ & $\frac{Q_S}{e\cdot N_\gamma^{sim}\cdot G} $\\
(atm) &(${}^\circ$K) && (fC)&&\\
\hline
  0.830 & 282.5 & $69\pm 5$ & $ 25.6\pm 1.0$ & 3238 & $ 0.72\pm 0.05$ \\
  0.777 & 281.5 & $67\pm 5$ &  $24.5\pm 0.9$ & 2956 &  $0.77\pm 0.06$ \\
  0.750 & 281.4 & $57\pm 4$ & $ 17.7\pm 0.7$ & 2815 & $ 0.69\pm 0.05$ \\
  0.719 & 281.3 & $57\pm 4$ &  $16.7\pm 0.5$ & 2659 & $ 0.69\pm 0.05$ \\
  0.678 & 281.2 & $53\pm 4$ & $ 15.5\pm 0.7$ & 2461 & $ 0.74\pm 0.06$ \\
\hline
Ar &DUV-enh & & & & $0.72\pm 0.06$ \\
\hline
\end{tabular}
\end{table}

The external quantum efficiency of APDs, from now on referred to as
quantum efficiency, is defined as the number of primary electron-hole
pairs produced per incident photon. The quantum efficiency
$\epsilon_Q(\lambda)$ is a function of the wavelength of the incident
light.

The quantum efficiency and APD gain $G$ relate to the charge signal $Q_S$ via 
\begin{equation}
G\cdot e\cdot \epsilon_Q(\lambda)\cdot N_\gamma^{sim}=Q_S
\label{eq:eqcal}
\end{equation}
where $e$ is just the elementary charge. By comparing $N_\gamma^{sim}$
with the measured charge signal, the quantum efficiency can be stated,
see Tables~\ref{tab:ar},\ref{tab:krxe},\&\ref{tab:xe}.

\begin{table}
\renewcommand{\arraystretch}{1.3}
\caption{\label{tab:krxe}The gain-independent signal normalized to the
expected number of VUV photons in krypton (See text). Measurements
performed using 'DUV-enhanced' LAAPD at 10$\mu$s shaping time. The last
line gives the average value of the measurements.}  
\centering
\begin{tabular}{|c|c|c|c|c|c|}
\hline
  $p_{Kr}$  & $T_{Kr}$  & Gain & $Q_s$  &  $N_\gamma ^{sim}$ & $\frac{Q_S}{e\cdot N_\gamma^{sim}\cdot G} $\\
(atm) &(${}^\circ$K) && (fC)&&\\
\hline
  0.629 & 285.0 & $26.1\pm 1.8$ &  $15.6\pm 0.6$ & 3844 &  $0.97\pm 0.08$ \\
  0.600 & 284.4 & $28.1\pm 2.1$ &  $15.8\pm 0.6$ & 3613 & $ 0.97\pm 0.08$  \\
  0.573 & 284.0 & $25.1\pm 1.6$ &  $15.5\pm 0.6$ & 3402 & $ 1.13\pm 0.08$  \\
  0.551 & 283.9 & $25.3\pm 1.6$ &  $14.3\pm 0.5$ & 3246 & $ 1.08\pm 0.08$  \\
  0.514 & 283.9 & $27.3\pm 1.9$ &  $13.8\pm 0.5$ & 2960 & $ 1.06\pm 0.08$  \\
\hline
Kr & DUV-enh & & & & $1.04\pm 0.08 $ \\
\hline
\end{tabular}
\end{table} 

\begin{table}
\renewcommand{\arraystretch}{1.3}
\caption{\label{tab:xe}The gain-independent signal normalized to the
expected number of VUV photons in xenon (See text). Measurements
performed using 'DUV-enhanced' LAAPD at 10$\mu$s shaping time. The last
line gives the average value of the measurements.}  
\centering
\begin{tabular}{|c|c|c|c|c|c|}
\hline
  $p_{Xe}$  & $T_{Xe}$  & Gain & $Q_s$  &  $N_\gamma ^{sim}$ & $\frac{Q_S}{e\cdot N_\gamma^{sim}\cdot G} $\\
(atm) &(${}^\circ$K) & & (fC)&&\\
\hline
 0.430 & 286.4 & $26.1\pm 1.8$ &  $20.6\pm 0.8$ & 3786 & $ 1.29\pm 0.10$ \\
 0.407 & 285.3 & $25.6\pm 1.7$ &  $19.1\pm 0.7$ & 3551 & $ 1.31\pm 0.10 $\\
 0.379 & 284.9 & $25.6\pm 1.7$ &  $18.0\pm 0.7$ & 3256 &  $1.35 \pm 0.10$\\
 0.372 & 286.0 & $24.6\pm 1.6$ &  $16.0\pm 0.6$ & 3163 &  $1.28\pm 0.10$ \\
\hline
Xe &DUV-enh & & & & $1.3\pm 0.1$ \\
\hline
\end{tabular}
\end{table} 

\begin{figure}
\centering
\includegraphics[width=2.5in]{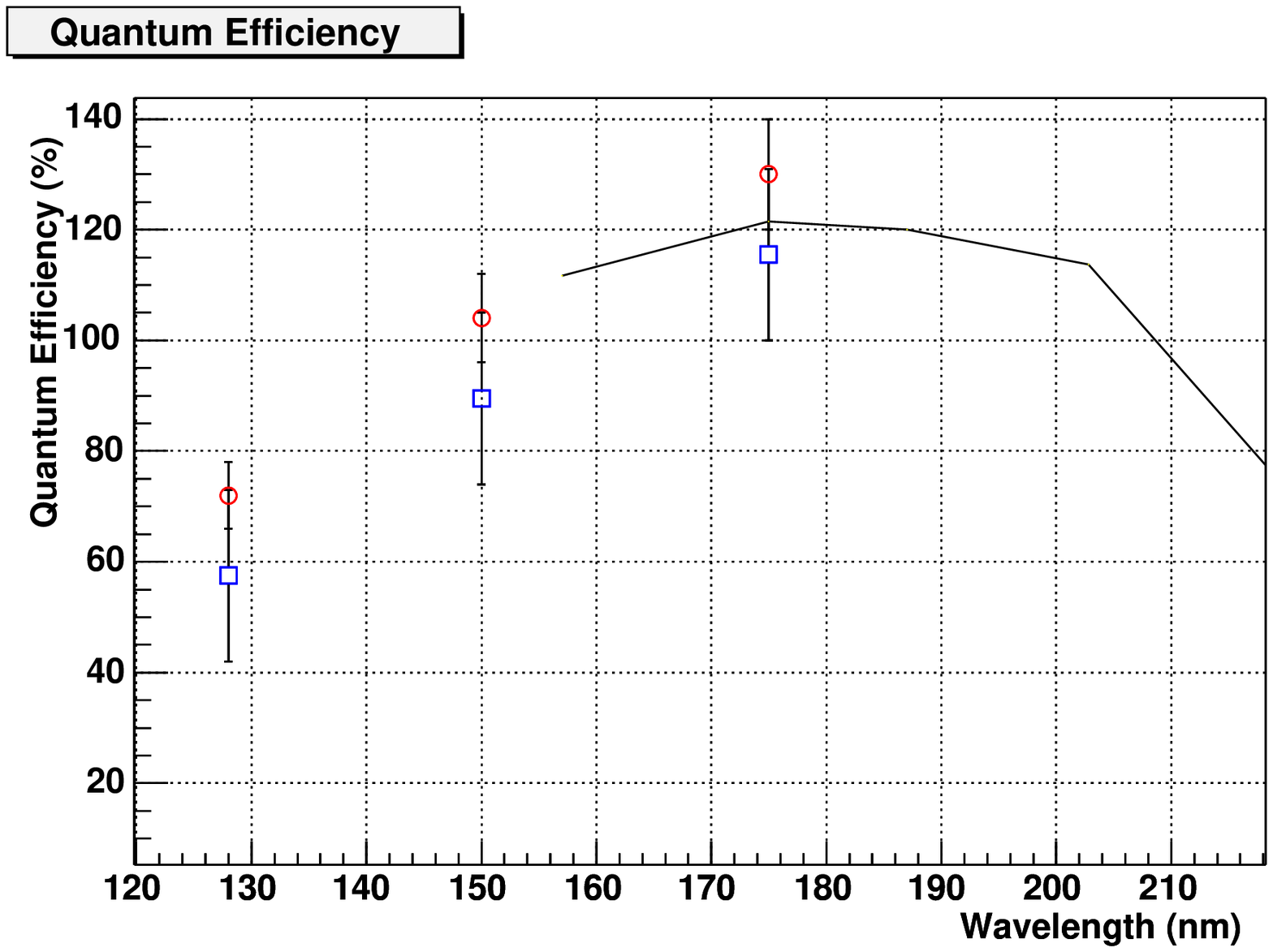} 
\caption{The quantum efficiency of the 'DUV-enhanced' LAAPD: The
continuous curve depicts the values given by the manufacturer
\cite{advphot}. The circles represent the values measured at argon,
krypton, and xenon emission wavelengths. The square markers include
the correction for non-UV components as measured in argon.}
\label{qeffdec}
\end{figure}

In the gas phase there is evidence that noble gas scintillation light
contains non-UV components. Additional measurements were performed to
subtract contributions from possible non-UV components of argon
scintillation light. This was done by employing a 'Red/IR-enhanced'
LAAPD of the same make. The device has a 150nm thick $SiO_2$
anti-reflective coating, rendering it largely insensitive to VUV
light. In a further measurement,a 0.1 mm thick UV-attenuator of
plastic foil was mounted in front of the 'Red/IR-enhanced' LAAPD, see
Table~\ref{tab:arir}.

A quantitative estimate of the non-UV emission components of argon
emission was achieved. An upper limit for the average amount of energy
needed to produce a non-UV photon in argon gas around at near ambient
pressure is found to be
\begin{equation}
 W_\gamma^{IR} \le 378 \, \rm  eV.
\end{equation}

The quantum efficiencies measured in
Tables~\ref{tab:ar},\ref{tab:krxe},\&\ref{tab:xe} need to be corrected
for the non-UV components (See Appendix). This was done in a conservative way,
leading to a lower limit of 42\% for the true quantum efficiency
128~nm. The non-UV components were only measured in argon, see Figure
\ref{qeffdec}.

\begin{table}
\renewcommand{\arraystretch}{1.3}
\caption{\label{tab:arir}Measurements performed in argon with the
'red/IR-enhanced' LAAPD. The middle line gives the average value of the
top three measurement series. The lower half of the table shows
results of measurements performed with a UV absorbing foil in front of
the 'red/IR-enhanced' LAAPD (See text). The average value of these
measurement series is given in the last line.}
\centering
\begin{tabular}{|c|c|c|c|c|c|}
\hline
 $p_{Ar}$  & $T_{Ar}$  & Gain & $Q_s$  &  $N_\gamma ^{sim}$ & $\frac{Q_S}{e\cdot N_\gamma^{sim}\cdot G} $\\
(atm) &(${}^\circ$K) && (fC)&&\\
\hline
  0.835 & 275.9 & 79.9 & 10.2  & 3388 &  0.23 \\
 0.820 & 276.7 & 77.6 & 9.39  & 3281 &  0.23 \\
 0.806 & 276.7 & 80.0 & 9.98  & 3197 &  0.24 \\
\hline
Ar& Red/IR-enh  &  &  &  & 0.233\\
\hline
0.862 & 278.7 & 90.4 & 7.2  & 3505 &  0.14 \\
0.847 & 279.2 & 159.9 & 13.6  & 3400 &  0.16 \\
0.830 & 279.1 & 168.5 & 14.5  & 3303 &  0.16 \\
\hline
Ar & Red/IR+Filter&&&& 0.153\\
\hline
\end{tabular}
\end{table} 

\begin{table}
\renewcommand{\arraystretch}{1.3}
\caption{\label{tab:res} Results of the quantum efficiency
measurements compared to the manufacturer's data where available.}
\centering
\begin{tabular}{|c|c|c|c|c|}
\hline
 $$  & $\lambda$ (nm) & QE $\ge$ & QE $\le$  &  QE manufacturer\\
\hline
  Ar & 128  & 42$\%$  & 73$\%$ & n.a. \\
  Kr & 150 & 74$\%$  & 113$\%$ & 112$\%$ @157nm \\
  Xe & 175 & 100$\%$  & 140$\%$ & 123$\%$  \\
\hline
\end{tabular}
\end{table} 

The obtained values agree well with
the data given by the manufacturer for xenon and krypton emission
wavelengths, see Figure~\ref{qeffdec}.

%

\section{Reflectivity Measurement}
\label{refl}

In our search for efficient light collection techniques, we included a
variable angle VUV reflector in our setup, see
Figure~\ref{fig:setup}. The reflector consists of a 125~$\mu$m Mylar
foil overcoated with 85~nm aluminum and 25~nm MgF$_2$ to prevent
aluminum oxidation.

The aluminum reflector evaporated on flexible
mylar foil is overcoated with a 20nm $MgF_2$ film to prevent
oxidation.

\begin{figure} 
\includegraphics[width=2.5in]{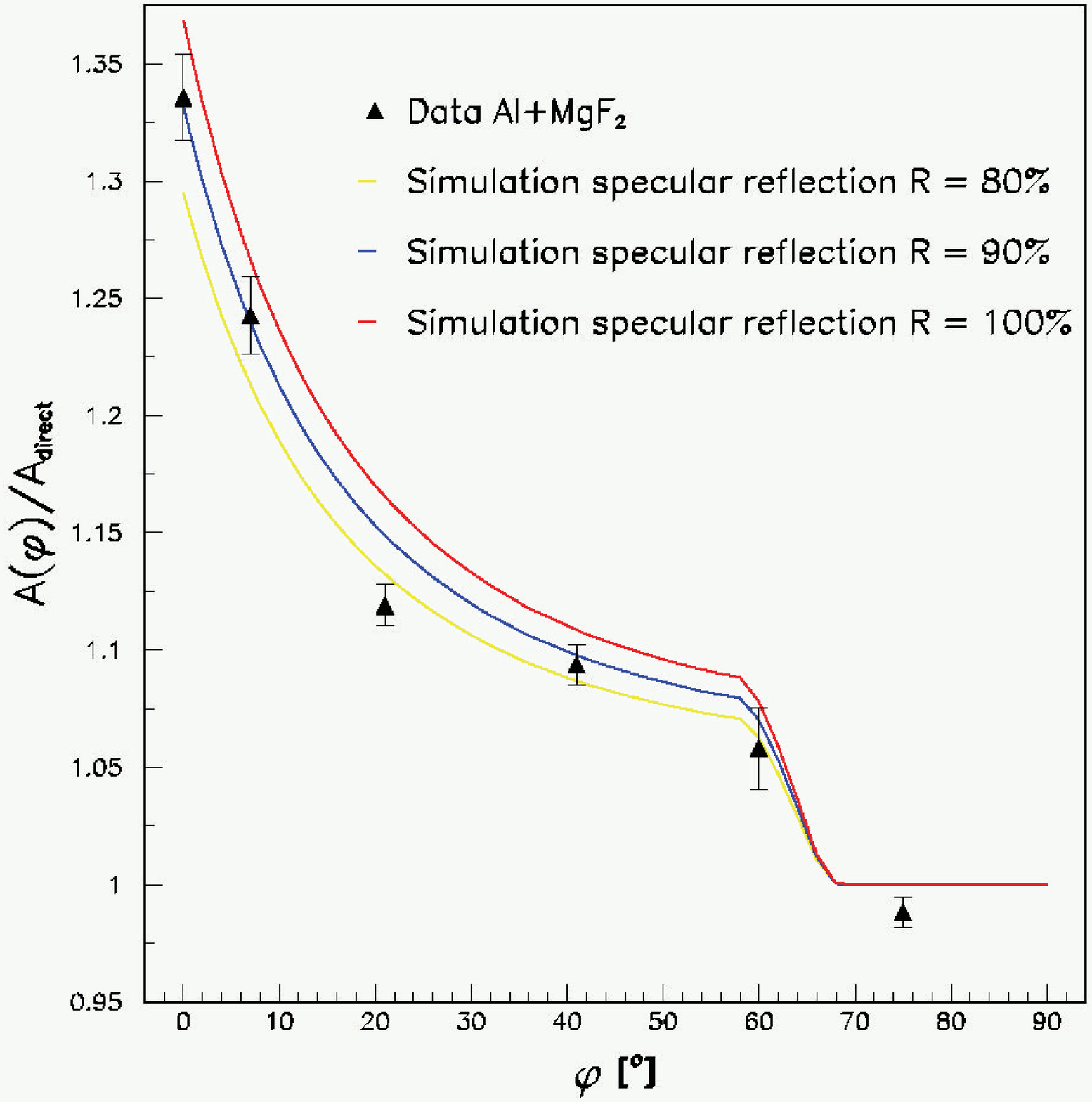} 
\centering
\caption{The angular dependence of the measured specular reflectivity of the Al+MgF$_2$ foil is
in good agreement with simulation for a specular reflectivity of \~90$\%$ at 128 nm.}
\label{arefl}
\end{figure}

Reflectivity measurements were performed by measuring the APD signal
height at different mirror angles. Each such measurement was normalized by
the signal height at $\phi=90$ degrees where no reflected light
contributes to the signal. These ratios were compared to the simulated
ratios $N_\gamma^{sim}(\phi)/N_\gamma^{sim}$, see
Figure~\ref{arefl}.

\begin{figure} 
\includegraphics[width=2.5in]{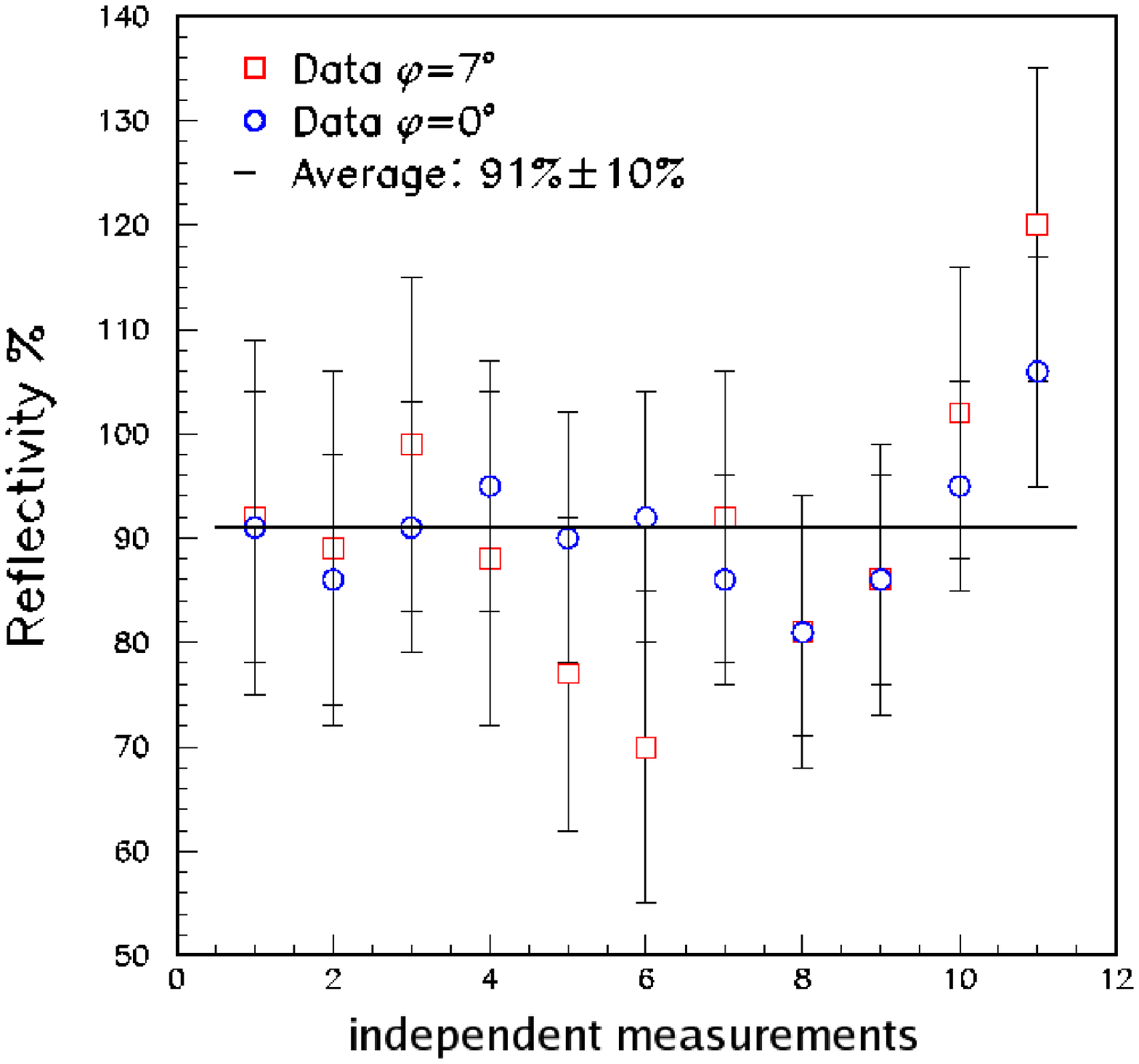} 
\centering
\caption{The measured specular reflectivity of the Al+MgF$_2$ foil is
in good agreement with the literature value of 90$\%$ at 128 nm.}
\label{fig:refl1}
\end{figure}

Our measured values are consistent with those of \cite{bradford},
stating a specular reflectivity of $\sim 90\%$ at 128 nm for
$Al+MgF_2$ on a glass substrate (See Figure~\ref{fig:refl1}).

\section{Conclusion}

Vacuum ultraviolet light can be detected by LAAPDs with improved
quantum efficiency compared to other means such as wavelength-shifter
coated photomultipliers. Figure~\ref{qeffdec} summarizes the obtained
results for argon, krypton, and xenon. Our measurements are consistent
with the manufacturer's data where it is available. The non-UV
correction was only measured for argon.  For xenon and krypton this
correction is of illustrative nature only. Since light quenching by remnant
impurities in the gas or by remnant air/water in the volume was not
considered, the lower limit of $\epsilon_q^{VUV}(128nm) \ge 42\%$ is
strict while the upper limit is not.

At room temperature, single photon counting was not achieved with the
device used. Preliminary tests suggest that even at low temperatures
the noise associated with this particular device's high capacitance
makes single photon detection difficult. This renders it impractical
for the use in ArDM. We do not exclude photon counting possibility
with smaller devices of lower capacitance from the same manufacturer,
operated at LAr temperatures.

The principle of VUV light collection by means of reflecting foils has
been shown to work. The fact that the foils are flexible makes them a
very handy tool for all liquid noble gas experiments. Applications
include the construction of Winston cones or other light concentrators.

\appendix 
\section*{Computation of $W_\gamma^{VUV}$ in Noble Gases}

The average energy expended per ion pair $W_g^{ion}$ can be related to
the ionization potential $I$ \cite{miyajima}
\begin{equation}
\label{eq:Wgi}
\frac{W_g^{ion}}{I}=\frac{E_i}{I} + \left(\frac{E_{ex}}{I}\right)\left(\frac{N_{ex}}{N_i}\right)+ \frac{\epsilon}{I},
\end{equation}
where $N_i$ is the number of ions produced at an average energy
expenditure of $E_i$, $N_{ex}$ is the number of excited atoms produced
at an average expenditure of $E_{ex}$, and $\epsilon$ is the average
kinetic energy of escape electrons.

Equation \ref{eq:Wgi} is energy dependent in all four terms, however,
for $E \gg I$ this dependence is weak. For $\alpha$-particles in
argon, $W_g^{ion} = 26.5\pm 0.5$~eV at energies $E_\alpha \ge
1$~MeV. For $E_\alpha = 0.1$~MeV, the value is only somewhat higher at
$W_g^{ion} = 27.5\pm 1.0$~eV, increasing further as the kinetic energy
is reduced. Our measurements were performed in a pressure range where
the scintillation is brought forth by $\alpha$ particles with at least
0.5~MeV kinetic energy.

Assuming no ionization contribution to UV scintillation light, justified at
the low pressures used in this work~\cite{suzuki}, the average energy expenditure
per photon is
\begin{equation}
\label{eq:Wg}
W_\gamma^{VUV} = \left(\frac{N_i}{N_{ex}}\right)\cdot E_i + E_{ex} +\left(\frac{N_i}{N_{ex}}\right)\cdot \epsilon
\end{equation}
electron volts. Substituting Equation~\ref{eq:Wgi} in
Equation~\ref{eq:Wg} gives
\begin{equation}
\label{eq:Wgtr}
W_\gamma^{VUV} = W_g^{ion}\cdot \frac{N_i}{N_{ex}}
\end{equation}
The values of $W_g^{ion}$ and $\frac{N_{ex}}{N_{i}}$ can be obtained
 from~\cite{platzman}, giving $W_\gamma^{VUV}=67.9$ eV, $61.2$ eV, and
 $55.9$ eV for Ar, Kr, and Xe, respectively.

\section*{Estimation of $W_\gamma^{IR}$}

Using the data listed in Table \ref{tab:arir}, we can give a strict
lower limit for the branching ratio $N^{IR}/N^{sim}_\gamma $ of the
emission of non-UV photons in argon. The number of non-UV photons
impinging on the 'Red/IR-enhanced' LAAPD relates to the detected charge
signal in linear dependence of gain and quantum efficiency:
\begin{equation}
N^{IR}=\frac{Q_S}{e\cdot \epsilon_Q(\lambda) \cdot G}
\end{equation}
The expression is minimized by the maximum quantum efficiency of the
Red/IR-enhanced APD, giving the expression
\begin{equation}
\label{eq:ir}
\left(\frac{Q_S}{e\cdot N_\gamma^{sim} \cdot
G}\right)_{Red/IR+Filter}\cdot \frac{1}{\max \epsilon_Q(\lambda)} = 0.18 \le \frac{N^{IR}}{N_\gamma^{sim}} ,
\end{equation}
where
\begin{equation}
  \max_{270\le\lambda\le1050} \epsilon_Q(\lambda)=0.85 
\end{equation}
was used. 

By comparison with $W_\gamma^{VUV}$ (See Equation~\ref{eq:Wgtr}), the
obtained value can be translated into
\begin{equation}
 W_\gamma^{IR} \le 378 \, \rm  eV,
\end{equation}
 a strict upper limit for the average amount of energy needed to
produce a non-UV photon in argon gas around this pressure.

\section*{Non-UV Correction for $\epsilon_Q(128nm)$}

With the data in Table \ref{tab:arir}, the quantum efficiency of the
'DUV-enhanced' LAAPD for radiation at 128 nm can be calculated more
precisely. In Equation \ref{eq:eqcal}, the measured charge signal
$Q_S$ (See Table \ref{tab:ar}) is actually the sum
\begin{equation}
Q_S=eG(N_\gamma \epsilon_Q(128nm) + N^{IR}(\lambda)\epsilon_Q(\lambda)) 
\end{equation}
of the VUV contribution and of the non-UV contribution, both weighted
with the quantum efficiency of the 'DUV-enhanced' APD at the respective
wavelength. The equation can be rewritten as
\begin{equation}
 \epsilon_{Q}^{uncorrected}= \frac{N_\gamma \epsilon_Q(128 nm) + N^{IR}(\lambda)\epsilon_Q(\lambda) }{N_\gamma}
\end{equation}
and
\begin{equation}
\epsilon_Q(128 nm) =  \epsilon_{Q}^{uncorrected}-  \frac{N^{IR}(\lambda)}{N_\gamma}\cdot\epsilon_Q(\lambda) .
\end{equation}

The value of the ratio $N^{IR}/N_\gamma^{sim}$ is contained in an
interval given by the measurements in Table~\ref{tab:arir} in a similar
way as was done in Equation~\ref{eq:ir}. Thus, a strict lower limit, 
\begin{eqnarray*}
\epsilon_Q(128nm) &\ge & \left(\frac{Q_S }{e\cdot N_\gamma^{sim}\cdot G}\right)_{VUV}\\
 & &    -\left(\frac{Q_S }{e\cdot N_\gamma^{sim}\cdot G}\right)_{IR}\cdot \max_{\lambda} 
\frac{\epsilon_q^{VUV}(\lambda)}{\epsilon_q^{IR}(\lambda)}
\end{eqnarray*}
and, as an upper limit
\begin{eqnarray*}
\epsilon_Q(128nm)& \le & \left(\frac{Q_S }{e\cdot N_\gamma^{sim}\cdot G}\right)_{VUV}\\
& &-\left(\frac{Q_S }{e\cdot N_\gamma^{sim}\cdot G}\right)_{IR+Filter}\cdot \min_{\lambda} 
\frac{\epsilon_Q^{VUV}(\lambda)}{\epsilon_q^{IR}(\lambda)}
\end{eqnarray*}
can be given, superindices denoting which APD the quantum efficiency
refers to, subindices labelling measurements from the right most
column of Tables~\ref{tab:ar}~and~\ref{tab:arir}. In the above
equations, the extrema of the ratio
$\epsilon_Q^{VUV}(\lambda)/\epsilon_Q^{IR}(\lambda)$ of the quantum
efficiencies of the two LAAPD types used can be obtained from the
manufacturer's data sheet on an interval from 270nm to 1050nm.
  
This gives 
\begin{equation}
0.42 \le \epsilon_q^{VUV}(128nm) \le 0.73
\end{equation}
where the upper limit is not rigorous as quenching effects due to gas
impurities have not been accounted for.

If argon non-UV emission is centered around 940 nm as measured by
\cite{bressi}, this would result in an in-between value of
$\epsilon_Q^{VUV}(128nm)\approx 0.58 $. Note that over a large region
of the IR spectrum, the ratio of $\epsilon_Q^{VUV}/\epsilon_q^{IR}$ is
relatively constant, making $\epsilon_Q^{VUV}(128nm)$ relatively
insensitive to the exact wavelength of peak IR emission. 

%
%


\section*{Acknowledgment}
This work is presented on behalf of the ArDM group, consisting of
scientists working for ETH Z\"urich and the University of Z\"urich in
Switzerland, CIEMAT and the University of Granada in Spain, and the
Soltan Institute Warsaw in Poland. We thank the Thin Film Workshop at
CERN, in particular A. Braem, for providing the reflecting foils and
useful discussions.




%




\end{document}